\documentclass[%
 reprint,
superscriptaddress,
 amsmath,amssymb,
 aps,
pra,
]{revtex4-1}
\usepackage{amsmath}
\usepackage{graphicx}
\usepackage{dcolumn}
\usepackage{bm}
\usepackage{hyperref}

\begin{document}

\preprint{APS/123-QED}

\title{Coherent post-ionization dynamics of molecules based on adiabatic strong-field approximation}

\author{Shan Xue}
 \affiliation{School of Nuclear Science and Technology, Lanzhou University, Lanzhou 730000, China}
  
\author{Wenli Yang}
\affiliation{School of Nuclear Science and Technology, Lanzhou University, Lanzhou 730000, China}

\author{Ping Li}
\affiliation{School of Nuclear Science and Technology, Lanzhou University, Lanzhou 730000, China}

\author{Yuxuan Zhang}
\affiliation{School of Nuclear Science and Technology, Lanzhou University, Lanzhou 730000, China}

\author{Pengji Ding}
\email{dingpj@lzu.edu.cn}
\affiliation{School of Nuclear Science and Technology, Lanzhou University, Lanzhou 730000, China}

\author{Song-Feng Zhao}
\affiliation{
 Key Laboratory of Atomic and Molecular Physics and Functional Materials of Gansu Province, College of Physics and Electronic Engineering, Northwest Normal University, Lanzhou 730070, China
}%

\author{Hongchuan Du}
\email{duhch@lzu.edu.cn}
\affiliation{School of Nuclear Science and Technology, Lanzhou University, Lanzhou 730000, China}

\author{Anh-Thu Le}
\affiliation{
 Department of Physics, University of Connecticut,  Storrs, Connecticut 06268, United States}%


\date{\today}

\begin{abstract}
Taking N$_2$ and O$_2$ as examples, we theoretically study the post-ionization dynamics of molecules in strong laser fields using a density matrix method for open systems, focusing on the effect of ionization-produced coherence on the dynamics of residual ions. We introduce the adiabatic strong-field approximation (ASFA) method to predict the coherences between ionic states resulting from multiorbital strong-field ionizations.  
Compared with the standard SFA, the ASFA method can be applied across a wide range of laser intensities due to the consideration of orbital distortion effects.
Based on the ASFA, it is found that there are obvious coherences between ionic states in residual molecular ions. These coherences significantly influence the transitions between ionic states, which finally change the post-ionization dynamics of molecular ions.  
Our findings reveal the importance of the ionization-produced coherences in modulating post-ionization molecular dynamics.

\end{abstract}

\maketitle


\section{INTRODUCTION}
When molecules are exposed to intense laser pulses, strong-field ionizations from multiple molecular orbitals (MOs) could create coherent superpositions of ionic states \cite{Brian2008hhg,smirnova2009high,Pabst2016}. These coherences can induce oscillatory charge migration, the frequency of which is determined by the energy gap between ionic states \cite{goulielmakis2010real,Goulielmakis2011,PhysRevLett.111.243005,calegari2014ultrafast}. Driven by electron-nuclear interactions, the electron motion subsequently triggers atomic rearrangements within the molecule on femtosecond to picosecond timescales \cite{breidbach2003migration,breidbach2005universal,belshaw2012observation,hennig2005electron,worner2017charge}. Revealing the mechanisms of coherence generation during strong-field ionization and its impact on the dynamics of the residual molecular ions is crucial for comprehending strong-field post-ionization phenomena, which ultimately paves the way for coherent control in biochemical reaction processes.

To theoretically investigate the post-ionization molecular dynamics, one should consider both the multi-electron effect and the nuclear motion. This makes the utilization of the full-dimensional time-dependent Schr$\ddot{\text{o}}$dinger equation (TDSE) impractical. So far, full-dimensional TDSE calculations for molecules have been constrained to small molecules like H$_2$ \cite{palacios2006enhancement,ishikawa2015review}. \textit{Ab-initio} calculations for more complex molecular systems remain scarce \cite{vacher2017electron}. 
To address this issue, an open-system density matrix (DM) method has been developed recently \cite{PhysRevA.100.031402,zhang2020sub,lei2022ultraviolet,yuen2022density,yuen2023modeling}. By incorporating the ionic reduced DMs (RDMs) generated by instantaneous strong-field ionizations of parent molecules into the DM equations (see Eq.~(\ref{eq:QLE})), the strong-field ionization, electric dipole coupling, and nuclear motion could be treated on equal footing in principle. To date, open-system DM methods primarily incorporate the ionic state populations generated by strong-field ionizations \cite{PhysRevA.100.031402,zhang2020sub,lei2022ultraviolet}. The off-diagonal terms, namely the ionization-produced coherences, have not been adequately addressed.
Despite this, theoretical simulations still qualitatively reproduce experimental results \cite{PhysRevA.100.031402,zhang2020sub,lei2022ultraviolet,zhang2020sub,lei2022ultraviolet}. 
In fact, laser ionizations can produce coherences in the residual ions. For atomic systems, the electronic coherences produced by photoionization and strong-field ionization have been studied \cite{rohringer2009multichannel,Pabst2016,carlstrom2017quantum}. 
But for molecules, the situation becomes more complicated due to nuclear motions. Several studies focus on the coherence evolution in polyatomic molecular ions produced by single-photon ionization. It has been found that the electronic coherence typically decays over a timescale of a few to hundreds of femtoseconds due to the dephasing and the decreased overlap of the nuclear wavepacket \cite{vacher2015electron,arnold2017electronic,vacher2017electron}. 
However, for molecules ionized by strong-field ionization, the generation of coherences within the residual ions has not been adequately studied. Moreover, the influence of the coherence on the dynamics of molecular ions remains unclear.

Based on the strong-field approximation (SFA), Pabst \textit{et al.}~proposed an intuitive approximative method that effectively predicted the coherence produced by strong-field ionization in xenon atoms \cite{Pabst2016}. Recently, we extended this method to diatomic molecular systems and verified its validity on a one-dimensional H$_2$ system \cite{xue2021vibronic}. 
This method has also been used to explain the abnormal ellipticity dependence of N$_2^+$ air lasing \cite{zhu2023influence}. In this work, we refer to it as the ``Simple" method. 
Recently, Yuen \textit{et al.}~developed an alternative approximative method based on partial wave expansion (PWE) \cite{PhysRevA.108.023123}. At the lowest order of the weak field asymptotic theory, the ionization amplitudes of partial waves are calculated, enabling the prediction of the ionic RDM \cite{tolstikhin2011theory}. This method achieves the same level of accuracy as the molecular Ammosov-Delone-Krainov (MO-ADK) theory in calculating the tunnelling ionization rate \cite{tong2002theory,Zhaosf2010}.
However, symmetry analysis suggests that ionizations from MOs with opposite parity should not generate coherence between the resulting ionic states. This contradicts the expectations of both the ``Simple" and the PWE methods, indicating that the generation of ionic coherences in molecules upon strong-field ionization remains elusive.

In this work, we investigate the ionic coherences generated by strong-field ionizations using the adiabatic strong-field approximation (ASFA), which is a combination of the SFA and the adiabatic field-distorted MOs. 
In Refs.~\cite{spiewanowski2013high,spiewanowski2014field,spiewanowski2015alignment}, it has been used to study the strong-field ionization and high-order harmonic generation of molecules. These works highlight the potential importance of orbital distortion in explaining experimental phenomena.
For molecular systems, coherences can be classified as vibrational coherence and vibronic coherence.
By using the ASFA, we find that the coherence between ionic states with opposite parities can be produced by strong-field ionization, which can not be captured by the standard SFA. 
Moreover, both of these coherences can influence the transitions between ionic states within the residual ions. Ultimately, the post-ionization dynamics of molecular ions are changed by these ionization-produced coherences.

The paper is structured as follows. In Sec.~II, we provide a brief review of the ``Simple" and PWE methods and introduce the (A)SFA method. We then present the open-system quantum Liouville equations.
In Sec.~III, using N$_2$ and O$_2$ as examples, we investigate the properties of ionization-produced coherences and their roles in post-ionization molecular dynamics. Section IV provides a summary. Atomic units are used throughout unless indicated otherwise.

\section{THEORETICAL METHODS}
For a molecule with $n$ electrons, the ionized molecular system can be described as an entangled state composed of the ion and the free electron at the instant of ionization. The state can be expressed as
\begin{equation} \label{eq:QLE}
\Psi(\mathbf{r_1},..,\mathbf{r}_{n})=\sum_{i,\mathbf{k}}c_{i}(\mathbf{k}) \hat{P}  _{mn}[\Phi_i(\mathbf{r_1},..\mathbf{r}_m,..\mathbf{r}_{n-1})\psi_{\mathbf{k}}(\mathbf{r}_{n}) ].  
\end{equation}%
Here, $\Phi_i$ represents the $i$-th ionic bound state generated by the ionization of the $i$-th MO in the neutral molecule. It is approximated using a Slater determinant. $\psi_{\mathbf{k}}$ represents the wavefunction of the resulting ionized electron with the momentum $\mathbf{k}$, approximated as a plane wave. $\hat{P }_{mn}$ is the antisymmetrizing permutation operator on the $m$-th and $n$-th electron coordinates.
$c_{i}(\mathbf{k})$ is the instantaneous ionization amplitude. 
Note that describing a molecular bound state using a single Slater determinant becomes inaccurate when electronic correlation effects are significant. To address this issue, one can adopt the density functional theory (DFT), which, in principle, incorporates both exchange and correlation effects. Alternatively, one can use more advanced post-HF methods, where the bound-state wavefunction is no longer restricted to a single Slater determinant.

In the basis set of ionic bound states, the ionic RDM produced by instantaneous ionization can be calculated by tracing out the free electrons, and is given by
\begin{equation}\label{eq:coherence}
 \rho_{ij}^{\text{ins}}(t)=\sum_{\mathbf{k}}c_i(\mathbf{k})c_j^*(\mathbf{k}).
 \end{equation}
The diagonal terms represent the ionization rates from the neutral state to different ionic states. The off-diagonal terms represent the instantaneous ionization-produced coherences.

\subsection{``Simple" method}
In the ``Simple" method, we denote the free electron wavepacket transiently ionized from the $i$-th molecular orbitals (MOs) as $\psi_{i}^{(\text{free})}$. It is assumed that the free-electron wavepackets ionized from different MOs share the same waveform \cite{Pabst2016,xue2021vibronic}, differing only by a complex coefficient $A_{i}=|A_{i}|e^{i\phi_i}$. This coefficient depends on the angle between the molecular axis and the field direction, the binding energy, and the wavefunction of the ionizing MO. Using a plane wave expansion, it means $\psi_{i}^{(\text{free})}(\mathbf{r})=A_i\int c(\mathbf{k})\psi_\mathbf{k}(\mathbf{r})d\mathbf{k}=\int c_i(\mathbf{k})\psi_\mathbf{k}(\mathbf{r})d\mathbf{k}$, where $c_i(\mathbf{k})=A_ic(\mathbf{k})$ is the ionization amplitude. 
Following Eq.~(\ref{eq:coherence}), the instantaneous ionization-produced RDM reads 
\begin{equation}\label{eq:Sins1}
\begin{aligned}
\rho_{ij}^{\text{ins}(\text{Simple})}&=\sum_{\mathbf{k}}c_i(\mathbf{k})c_j^*(\mathbf{k})  \\
&=\sum_{\mathbf{k}}|A_{i}|e^{\text{i}\phi_{i} }c(\mathbf{k})|A_{j}|e^{-\text{i}\phi_{j} }c^*(\mathbf{k})\\
&=\sum_{\mathbf{k}}|c_i(\mathbf{k})||c_j(\mathbf{k})|e^{\text{i}(\phi_{i}-\phi_{j}) }\\
&=  \sqrt{ \sum_{\mathbf{k}}|c_i(\mathbf{k})|^2\times\sum_{\mathbf{k}}|c_j^*(\mathbf{k})|^2}\times e^{\text{i}(\phi_{i}-\phi_{j})  }.
\end{aligned}
\end{equation}%
In the final step of the derivation, we utilize the equality condition in the Cauchy-Schwarz inequality, because $|c_i(\mathbf{k})|$ and $|c_j(\mathbf{k})|$ are linearly dependent. Strictly speaking, the assumption that $\psi_{i}^{(\text{free})}$ and $\psi_{j}^{(\text{free})}$ share the same waveform is too restrictive, and $|c_i(\mathbf{k})|/|c_j(\mathbf{k})|$ should depend on $\mathbf{k}$. According to the Cauchy-Schwarz inequality, the exact value of $|\rho_{ij}^{\text{ins}}|$ should be smaller than $|\rho_{ij}^{\text{ins}(\text{Simple})}|$. 
This indicates that the off-diagonal elements, namely the instantaneous ionization-produced coherences, calculated using the ``Simple" method, are somewhat overestimated.

In the following, we determine the phase difference $\phi_{i}-\phi_{j}$ through a simple derivation.
For homonuclear diatomic molecules, the MO wavefunction $\varphi_i$ has a definite even ($g$) or odd ($u$) parity. For the $g(u)$ orbital, the transition dipole moment (TDM) $\mathbf{u}_i(\mathbf{k})=-\left \langle \psi_\mathbf{k}|\mathbf{r}|\varphi_i \right \rangle $ is a $u(g)$ function of $\mathbf{k}$. Given that the instantaneous ionization amplitude $c_{i}(\mathbf{k},\mathbf{F})\propto  \mathbf{u}_{i}(\mathbf{k})\cdot \mathbf{F}$ ($\mathbf{F}=F\mathbf{e}_z$ is the electric field along the space $z$ axis), and strong-field ionization primarily occurs in the opposite direction of the electric field, one obtains $c_{i}(\mathbf{k},F)=c_{i}(-\mathbf{k},-F)$ for a $g$ MO and $c_{i}(\mathbf{k},F)=-c_{i}(-\mathbf{k},-F)$ for a $u$ MO. Therefore, when two ionizing MOs possess the same parity, $\rho_{ij}^{\text{ins(Simple)}}(F)=\rho_{ij}^{\text{ins(Simple)}}(-F)$. When they possess opposite parity, $\rho_{ij}^{\text{ins(Simple)}}(F)=-\rho_{ij}^{\text{ins(Simple)}}(-F)$. Hence, the exponential term in Eq.~(\ref{eq:Sins1}) can be further written as $e^{\text{i}(\phi_{i}-\phi_{j}) }=\text{sgn}[F]^{(2-P_i-P_j)/2}$. Here, $\text{sgn}$ is the sign function. $P_{i(j)}$ describes the parity of the $i(j)$-th ionizing MO, with a value of $+1(-1)$ for $g(u)$ parity.  
By replacing $\sum_{\mathbf{k}}|c_i(\mathbf{k})|^2$ with the MO-ADK ionization rate $\Gamma_i(F(t))$ \cite{tong2002theory,Zhaosf2010}, Eq.~(\ref{eq:Sins1}) can be rewritten as
 \begin{equation} \label{eq:Sins2}
 \rho_{ij}^{\text{ins(Simple)}}(t)=\sqrt{\Gamma_i(t)\Gamma_j(t)}\text{sgn}[F(t)]^{(2-P_i-P_j)/2}.
\end{equation}%
Considering the free evolution of the ionization-produced RDM, the RDM at a final time $t_f$ can be given by
\begin{small}
\begin{equation} 
\begin{aligned}
&\rho_{ij}^{\text{(Simple)}}(t_f)
\\\!=&\!\!\!\int_{-\infty}^{t_f}\!\!dt\!\sqrt{\Gamma_i(t)\Gamma_j(t)}\text{sgn}[F(t)]^{(2-P_i-P_j)/2}e^{i(E_i-E_j)t}\!e^{-i(E_i-E_j)t_f}.
\end{aligned}
\end{equation} 
\end{small}%
The same equation can also be found in Ref.~\cite{Pabst2016}, differing only by the constant term $e^{-i(E_i-E_j)t_f}$, which has no substantial impact on the results.

\subsection{Partial-wave-expansion method}
Based on the weak-field asymptotic theory and PWE  \cite{tolstikhin2011theory,SONG2023108882}, Yuen \textit{et al.} proposed an alternative method to calculate the ionization-produced RDM \cite{PhysRevA.108.023123},
\begin{equation}  
\rho_{ij}^{\text{ins(PWE)}}=\sum_m \gamma_{im}(F)\gamma_{jm}^*(F). 
\end{equation}%
Here, $\gamma_{im}$ is the partial ionization amplitude from the $i$-th ionizing orbital. $m$ is the magnetic quantum number along the space $z$ axis. Using the adiabatic approximation, $\gamma_{im}$ is expressed as
\begin{equation} \label{eq:PWE2}
\begin{aligned}
\gamma_{im}&=\frac{B_{im}}{\sqrt{2^{|m|}|m|!}} \frac{1}{\kappa_i^{Z/\kappa_i-1/2} } 
\left(   \frac{2\kappa_i^3 }{|F(t)|}  \right)^{Z/\kappa_i-(|m|+1)/2}\\
&\times \text{exp}\left[ \frac{-\kappa_i^3 }{3|F(t)|} +\frac{\text{i}\pi}{4}+\text{i}\pi \left( \frac{Z}{\kappa_i}-\frac{|m|+1}{2}   \right) \right],
\end{aligned}
\end{equation}%
where $\kappa_i=\sqrt{2I_{pi}}$ with $I_{pi}$ being the ionization potential of the $i$-th ionizing orbital. $Z=1$ is the effective asymptotic charge experienced by the ionized electron. Considering that the electron is ionized in the opposite direction of the electric field $F$, $B_{im}$ can be given by
\begin{small}
\begin{equation}  
\begin{aligned}
B_{im}\!\!=\!\!\begin{cases}
\sum_{l,m_0} C_{i,lm_0}D^l_{mm_0}(\alpha,\beta,\gamma)Q(l,m)~~~~~~~~~~~~~~F<0  \\
\sum_{l,m_0} (-1)^{l-m_0}C_{i,lm_0}D^l_{mm_0}(\alpha,\beta,\gamma)Q(l,m)~~F>0, 
\end{cases}
\end{aligned}
\end{equation}%
\end{small}%
where 
\begin{equation}  
Q(l,m)\!\!=\!\!(-1)^{(m+|m|)/2} \sqrt{(2l+1)(l+|m|!)/2(l-|m|)!}.
\end{equation}%
Here, $C_{i,lm_0}$ is the structure parameter of the $i$-th ionizing orbital \cite{Zhaosf2010}. $l$ is the orbital angular momentum quantum number. $m_0$ is the magnetic quantum number along the molecular axis. $D^l_{mm_0}$ represents the Wigner D-matrix for rotating the molecule.   
Notably, this method aligns with the MO-ADK theory for calculating ionization rates, which allows for the computation of tunnelling ionization rates from MOs at different molecular axis angles.

\subsection{Adiabatic strong-field approximation method}
In this section, we introduce the ASFA coherence method.
In the length-gauge SFA, the amplitude for producing an ion in the $i$-th state, accompanied by a free electron with the momentum $\mathbf{p} $ at the end of the laser field $t_f$, is given by \cite{Pabst2016,spiewanowski2014field}
\begin{equation} \label{eq:ASFA1}
M_i(\mathbf{p},t_f)=-\text{i}\int^{t_f}_{-\infty } \mathbf{u}_{i}(\mathbf{k}(t))\cdot\mathbf{F}(t)e^{-\text{i}\int_{t}^{t_f}[|\mathbf{k}(t')|^2/2-E_{i}(t')]dt'}dt.
\end{equation}%
Here, $\mathbf{u}_{i}(\mathbf{k})=-\left \langle  \psi_{\mathbf{k}}|\mathbf{r}|\varphi_i(t)  \right \rangle $ is the TDM element from the ionizing MO $\varphi_i(t)$ to the plane wave $\psi_{\mathbf{k}}$. $\mathbf{k}(t)=\mathbf{p}-\mathbf{A}(t_f)+\mathbf{A}(t)$ with $\mathbf{A}$ being the vector potential of the laser field. $E_i(t)$ is the time-dependent eigenenergy of the $i$-th MO.
In the adiabatic approximation, the laser field at the ionization instant can be treated as a static electric field. This treatment results in the distortions of MOs. 
Then the eigenequation for the MO becomes 
\begin{equation} \label{eq:ASFA2}
\hat{h}(t)|\varphi_i(t) \rangle =E_i(t)|\varphi_i(t) \rangle.
\end{equation}%
Here, $\hat{h}(t)$ is the effective one-electron Hamiltonian operator, which is composed of the field-free Hamiltonian $\hat{h}_0$ and the interaction term $\mathbf{F}(t)\cdot \mathbf{r}$. In this work, adiabatic MOs are computed using the DFT as implemented in the GAUSSIAN16 program
package \cite{g16}. We adopted the B3LYP exchange-correlation functional and the augmented correlation-consistent polarized valence quadruple-zeta (aug-cc-pVTZ) basis set. The basis set incorporates diffuse functions, which can effectively characterize the field-induced diffusion of MOs.   

By tracing out the degrees of freedom of the ionized electron, we obtain the RDM of the ion
\begin{equation} \label{eq:ASFA3}
\rho_{ij}(t_f)=\int d\mathbf{p} M_i(\mathbf{p},t_f)M_j^*(\mathbf{p},t_f).
\end{equation}%
Substituting Eq.~(\ref{eq:ASFA1}) into the equation above and assuming that electron wavepackets with the same $\mathbf{p}$ but emitted at different times have negligible overlap due to rapid wavepacket spreading, this allows us to determine the coherence by neglecting the terms where two ionic states are populated at different times \cite{Pabst2016}. As a result, Eq.~(\ref{eq:ASFA3}) can be simplified as
\begin{subequations} \label{eq:ASFA4}
\begin{align}
\rho_{ij}(t_f)&=\int_{-\infty}^{t_f} \rho_{ij}^{\text{ins}}(t)e^{\text{i}\int_t^{t_f}[E_i(t')-E_j(t')]dt'}dt, \\
\rho_{ij}^{\text{ins}}(t)&= \!\!\int \left[\mathbf{u}_{i}(\mathbf{k}) \!\cdot\!\mathbf{F}(t)\!\right]\! \left[\mathbf{u}_{j}^*(\mathbf{k}) \!\cdot\! \mathbf{F}(t)\right]d\mathbf{k} .
\end{align}
\end{subequations}%
Here, $\rho_{ij}(t_f)$ is the RDM of the ion at the end of the laser pulse, when dipole transitions between ionic states are excluded. $\rho_{ij}^{\text{ins}}(t)$ represents the instantaneous ionic RDM generated at time $t$. 
To consider dipole transitions between ionic states, we can incorporate $\rho_{ij}^{\text{ins}}(t)$ into the open-system DM equation that describes the evolution of the ion system. 
To assess the degree of coherence (DOC) of the ion generated by instantaneous ionization, we define
\begin{equation} \label{eq:CDOC}
G_{ij}^{\text{(A)}}(t)=\rho_{ij}^{\text{ins}}(t)/\sqrt{\rho_{ii}^{\text{ins}}(t)\rho_{jj}^{\text{ins}}(t)} 
\end{equation}%
as the complex DOC. It contains both the DOC and the phase information of the coherence. Since $G_{ij}^{\text{(A)}}$ can be obtained using either SFA or ASFA, the superscript ``A" refers to either ``SFA" or ``ASFA."

In practical calculations, we replace the diagonal elements $\rho_{ii}^{\text{ins}}(t)$ with the MO-ADK ionization rate $\Gamma_i(t)$. To ensure that the DOC generated by instantaneous ionization is consistent with Eq.~(\ref{eq:CDOC}), we replace the coherence term $\rho_{ij}^{\text{ins}}(t)$ with $\sqrt{\Gamma_i \Gamma_j} G_{ij}^{\text{(A)}}$. Finally, the ionic DM generated by instantaneous ionization is denoted as $\rho_{ij}^{\text{ins(A)}}(t)$, with ``A"=``SFA" or ``ASFA".
In the following analysis, the coherence and ionic DM are also calculated by the ``Simple" and PWE coherence methods, where the subscript ``A" is replaced by ``Simple" and ``PWE", respectively.

\subsection{Quantum Liouville equations} 
Considering that the interaction between molecules and laser fields only lasts for a few tens of femtoseconds, the molecular rotation can be safely neglected. 
Consequently, the ionic density operator can be expanded in terms of vibronic states under the Born-Oppenheimer~(BO) approximation, and given by
$\rho_{vv'}^{ij}=\langle\Phi _i \chi_v^i |\hat{\rho}|\Phi _j \chi_{v'}^j \rangle$. Here, $\Phi _{i}$ and $\chi_v^i$ represent the $i$-th electronic state and $v$-th vibrational state on the $i$-th ionic potential energy curve, respectively. Accordingly, the open-system DM equations are formulated using the quantum Liouville equations
\begin{small}
\begin{equation} \label{eq:QLE}
\begin{aligned}
&\text{i}\frac{\partial}{\partial t}\rho_{vv'}^{ij}\\
=&\omega_{vv'}^{ij}\rho_{vv'}^{ij}-\mathbf{F}^{\text{(MF)}}\cdot\sum_{o,v''}(\mathbf{u}_{vv''}^{io}\rho_{v''v'}^{oj}-\rho_{vv''}^{io}\mathbf{u}_{v''v'}^{oj})+\text{i}\rho_{vv'}^{ij,\text{ins(A)}}, 
\end{aligned}
\end{equation}%
\end{small}%
where $\omega_{vv'}^{ij}=E_{v}^{i}-E_{v'}^{j}$ is the energy difference between vibronic states. $\mathbf{u}_{vv'}^{ij}=\langle\chi_v^i(R)|\mathbf{u}_{ij}(R)|\chi_{v'}^j(R) \rangle$ is the vibronic-state TDM with $\mathbf{u}_{ij}(R)=-\langle \psi_i(\mathbf{r},R)|\mathbf{r}| \psi_j(\mathbf{r},R) \rangle $ being the $R$-dependent electronic-state TDM. $\mathbf{F}^{\text{(MF)}}$ is the electric field in the molecule-fixed (MF) coordinates, with the $z$-axis along the molecular axis. $\mathbf{F}^{\text{(MF)}}$ can be transformed from the electric field in the space-fixed (SF) coordinates $\mathbf{F}^{\text{(SF)}}$ using $\mathbf{F}^{\text{(MF)}} = \mathbf{R}(\alpha, \beta,\gamma) \mathbf{F}^{\text{(SF)}}$ \cite{zhu2023influence}. Here, $\mathbf{R}(\alpha, \beta,\gamma)$ is the rotational matrix, with $\alpha$, $\beta$, and $\gamma$ denoting the Euler angles in the $zyz$ convention.

$\rho_{vv'}^{ij,\text{ins(A)}}$ represents the vibronic-state-resolved RDM of the ion produced by instantaneous ionization. By modelling the vibrational state distribution of the ion using the Franck-Condon (FC) factors $C_{v}^{i(\text{FC})}$, $\rho_{vv'}^{ij,\text{ins(A)}}$ can be calculated by
\begin{equation} \label{eq:ins}
\rho_{vv'}^{ij,\text{ins(A)}}(t)= \rho_{ij}^{\text{ins(A)}}\sqrt{C_{v}^{i(\text{FC})}C_{v'}^{j(\text{FC})}}\times n(t).
\end{equation}%
Here, $C_{v}^{i(\text{FC})}=|\langle \chi_v^i | \chi_0^{(\text{neutral})} \rangle|^2$ with $\chi_0^{(\text{neutral})}$ representing the ground vibronic state of the neutral molecule. $n(t)$ is the remaining probability of the neutral molecule at time $t$. $\rho_{ij}^{\text{ins(A)}}$ is the instantaneous ionization-produced ionic RDM calculated at the equilibrium internuclear distance of the neutral molecule. 
When $i = j$ and $v \neq v'$, $\rho_{vv'}^{ii,\text{ins(A)}}$ represents the instantaneous ionization-produced vibrational coherence (VC). When $i \neq j$ and $v \neq v'$, $\rho_{vv'}^{ij,\text{ins(A)}}$ represents the instantaneous ionization-produced vibronic coherence (VEC). In Sec.~III, we will investigate the effects of these two types of coherences on the post-ionization dynamics of molecular ions.
  
In Sec.~III~D, we investigate the dissociative ionization dynamics of O$_2$. Following single-electron ionization, the dissociation of O$_2^+$ produces O and O$^+$. The kinetic energy release (KER) spectrum is calculated by 
\begin{equation} \label{eq:KER}
C^{\text{KER}}(E_k)=\sum_i\rho^{ii(\text{dis})}_{vv}(E_k)/\sqrt{2\mu E_k }.
\end{equation}%
Here, $E_k$ is the dissociative energy. $\rho^{ii(\text{dis})}_{vv}(E_k)$ is the population of the $v-$th quasi-continuum state on the $i$-th potential energy curve of the ion at the end of the laser pulse. $\mu$ represents the reduced mass of the nuclei. The upper limit of the dissociation energy is set to 3 eV to capture most dissociation signals under current laser parameters.

\begin{figure} [htb]
\includegraphics[width=9cm,clip=true]{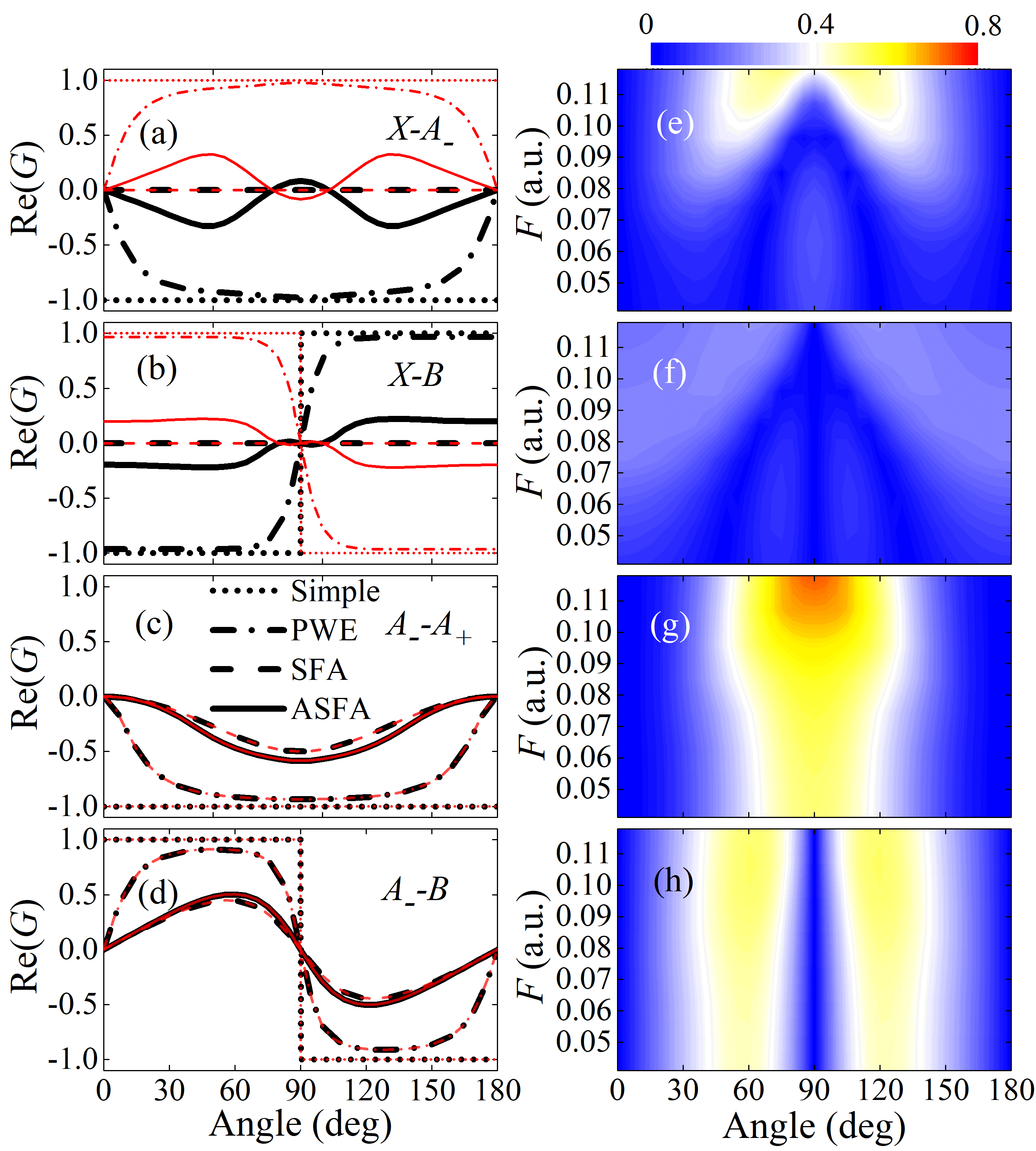}
\caption{(Color online) (a)-(d) Real parts of the complex DOCs, Re$(G_{ij})$, calculated using Eq.~(\ref{eq:CDOC}) at fixed field strengths of $F=0.09$~a.u.~(black thick lines) and $F=-0.09$~a.u.~(red thin lines). (e)-(h) DOCs, $|G_{ij}|$, as functions of field strength and molecular axis angle, calculated using the ASFA method.}
\label{figure11}
\end{figure}

\section{RESULTS AND DISCUSSION}
Taking N$_2$ and O$_2$ as examples, we investigate the influence of ionization-produced ionic coherence on the dynamics of residual ions after strong-field ionization.  
For N$_2$, we focus on the establishment of population inversion between states $X^2 \Sigma _g^+$ and $B^2 \Sigma _u^+$, which is closely related to the generation of N$_2^+$ lasing \cite{PhysRevA.84.051802,Xu2015,PhysRevLett.115.133203,PhysRevLett.119.203205,PhysRevLett.120.133208,PhysRevLett.129.123002}.
For O$_2$, we mainly discuss the dissociation process of O$_2^+$.

\subsection{Ionization-produced coherence in N$_2^+$}
When an intense laser pulse interacts with N$_2$, the strong-field single ionization primarily produces four ionic states: $X^2 \Sigma _g^+ $, $A^2 \Pi _{u\mp}$, and $B^2 \Sigma _u^+$, which come from the ionizing MOs $3\sigma_g$ (highest occupied molecular orbital (HOMO)), $1\pi_{u\pm}$ (HOMO-1), and $2\sigma_u$ (HOMO-2) in N$_2$, respectively. For simplicity, these states are abbreviated as $X$, $A\mp$, and $B$ in the following. $\pm$ on the subscript denotes the orbital angular momentum along the molecular axis, with values of $\pm 1$. 
Additionally, it is worth noting that the $X-B$ TDM is parallel to the molecular axis, while the $X-A\mp$ TDMs are perpendicular to the molecular axis \cite{zhu2023influence}.

To investigate the effect of ionization-produced coherence on the dynamics of N$_2^+$, we first examine the properties of the coherence between ionic states produced at fixed electric field strengths.
Figures~\ref{figure11}(a)-\ref{figure11}(d) show the complex DOC, $G_{ij}$, as a function of the angle between the molecular axis and the electric field, calculated using different coherence methods. Since only the PWE method gives the coherence with negligible imaginary parts, and the coherences calculated by other methods are purely real numbers (see Appendix),
we display only the real parts of $G_{ij}$. Several features can be observed. (i) Except for the SFA method, all other methods predict the existence of the $X-A_-$ and $X-B$ coherences. (ii) Although the magnitudes of Re($G_{ij}$) obtained by different methods are different, their signs are the same except for the results of ASFA at the angle of $\sim 90^\circ$. The structure of Re($G_{XA_-}^{(\text{ASFA})}$) around $90^\circ$ can be attributed to the counterintuitive shape of the distorted ionizing orbital $3\sigma_g$ (see Figs.~\ref{figure88}(a) and \ref{figure88}(c) in the Appendix). Moreover, when the electric field is reversed, the following relationships are satisfied:
\begin{equation}
\begin{aligned}
G_{XA_-}(F)&=-G_{XA_-}(-F),\\ 
G_{XB}(F)&=-G_{XB}(-F),\\
G_{A_-A_+}(F)&=G_{A_-A_+}(-F),\\
G_{A_-B}(F)&=G_{A_-B}(-F).
\end{aligned}
\nonumber
\end{equation}%
(iii) Quantitatively, the ``Simple" method predicts the highest DOC $|G_{ij}|$ of 100\%, followed by the PWE method. In comparison, the ASFA and SFA methods predict lower DOCs $|G_{ij}|$. 
In the ``Simple" method, we assume that the ionization primarily occurs in the opposite direction of the electric field. And the free-electron wavepackets ionized from different MOs are correlated with the same free-electron wavepacket. This means that the ionization amplitudes $c_i(\mathbf{k})$ and $c_j(\mathbf{k})$ are linearly dependent. As a result, the DOC $|G_{ij}|$ between the ionic states is 100\%, see Eq.~\ref{eq:Sins1}.
In the PWE method, ionization also occurs in the opposite direction of the electric field. But the partial ionization amplitudes $\gamma_{im}$ and $\gamma_{jm}$ given by Eq.~\ref{eq:PWE2} are not linearly dependent. Therefore, the DOC $|G_{ij}|$ is lower.
In the ASFA and SFA methods, $c_i(\mathbf{k})$ and $c_j(\mathbf{k})$ are still not linearly dependent. Moreover, ionization occurs in all directions. More ionization directions result in a lower probability of the generated ionic states being correlated with the same free electron. After tracing out the free electrons, the DOC is significantly reduced.

We further discuss the dependence of ionization-produced coherence on the field strength. According to Eq.~(\ref{eq:Sins2}), the DOCs predicted by the ``Simple" method do not rely on field strength. The same conclusion holds for the SFA method since the distortion of the ionizing MOs is not considered. For the PWE method, because the phase in Eq.~(\ref{eq:PWE2}) does not depend on the field strength, its predicted DOCs exhibit anomalous subtle decreases as the field strength increases (less than 5\% within $F\!\in\! 0.04\sim0.12$~a.u.).
Therefore, in Figs.~\ref{figure11}(e)-\ref{figure11}(h), we solely present the DOCs calculated by the ASFA method.  
As shown in Figs.~\ref{figure11}(g) and \ref{figure11}(h), the DOCs between ionic states with the same parity ( $A_--A_+$ and $A_--B$) slightly decrease with the decreasing field strength, but do not disappear. This indicates that these DOCs are not sensitive to orbital distortion. However, for ionic states with opposite parity ($X-A_{\pm}$ and $X-B$), the ionization-produced DOCs gradually emerge as the field strength increases, as shown in Figs.~\ref{figure11}(e) and \ref{figure11}(f). Therefore, these DOCs are more dependent on the orbital distortion and more sensitive to the instantaneous field strength.

For the above results of $X-A_-$ and $X-B$, we can provide the following explanation. At low field strengths, the orbital distortion is relatively small, and the parity of MOs can be well maintained.
For ionizing MOs with opposite parity, the corresponding photoelectron wavepackets also carry opposite parity, and maintain orthogonality with one another. In this case, the generated ionic states cannot correlate with the same electronic continuum state, so there is no coherence between the two ionic states. Thus, the $X-A_{\pm}$ and $X-B$ DOCs approach zero at weak field strengths, as shown in Figs.~\ref{figure11}(e) and \ref{figure11}(f). This implies that, within the multiphoton regime, the ionization-produced coherence primarily depends on the parities of the field-free MOs. On this condition, the SFA and ASFA methods give similar results.
As the laser intensity increases, the parities of MOs will be disrupted due to the distortion of MOs. The electron wavepackets ionized from the distorted MOs are no longer orthogonal, so that different ionic states may correlate with the same electronic continuum state.  
Consequently, the coherence between the ionic states emerges after tracing out the free electrons. Therefore, in the tunnelling regime, the distortions of the ionizing MOs are crucial for the calculation of ionization-produced coherence, which is not captured in the standard SFA.   

\begin{figure} [htb]
\includegraphics[width=9.2cm,clip=true]{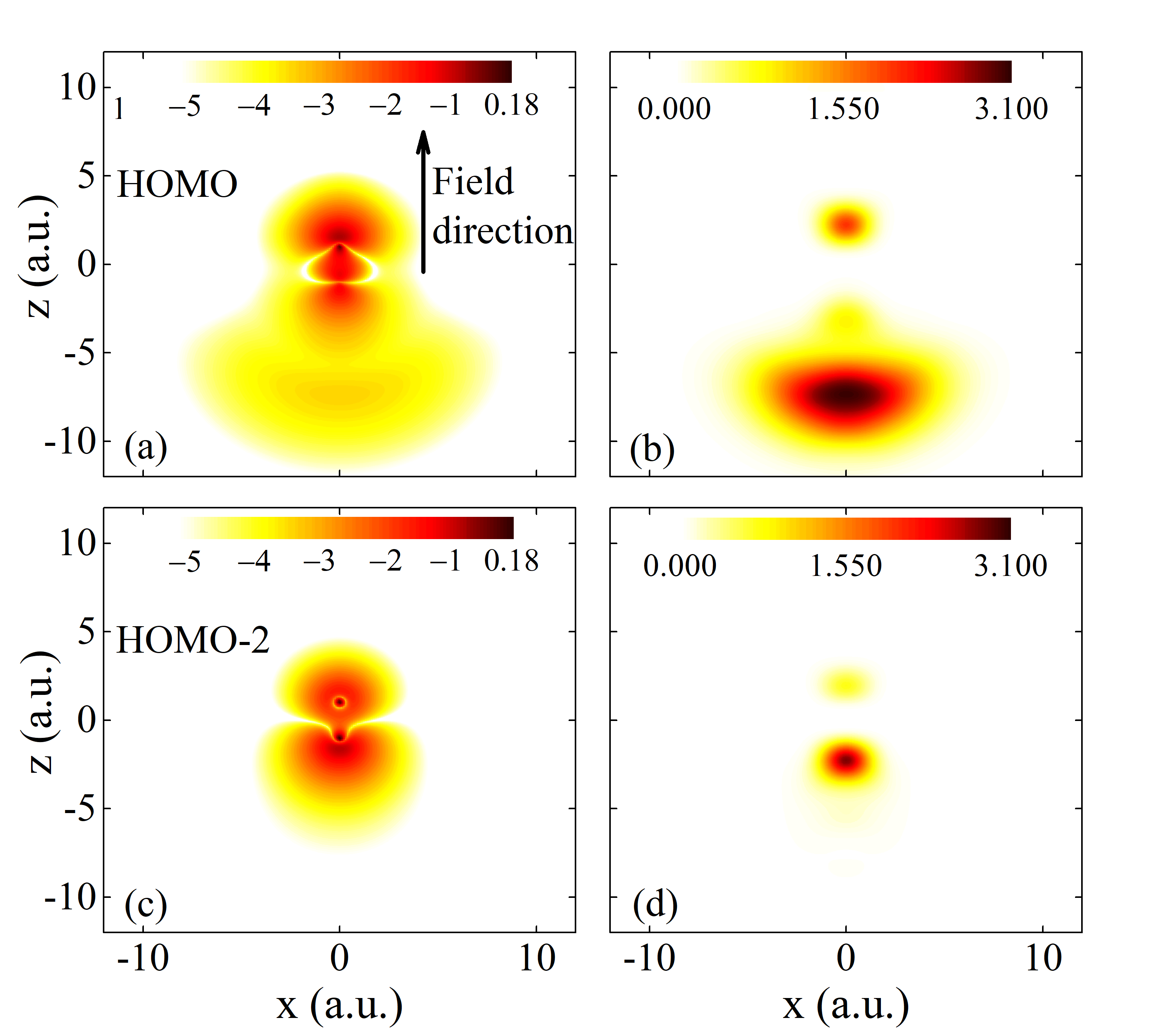}
\caption{(Color online) (a) Electron density in the $xoz$ plane of the field-distorted HOMO ($3\sigma_g$) at a fixed field strength of $F=0.08$~a.u. The molecular axis is parallel to the field direction. For clarity, the density is plotted on a logarithmic ($log_{10}$) color scale. (b) Density of the ionized electron at the ionization instant. (c) and (d) are the same as (a) and (b), but for HOMO-2 ($2\sigma_u$).}
\label{figure22}
\end{figure}

Fundamentally, only the free-field ground state is considered in the standard SFA. The lack of coupling with other states prevents MOs from exhibiting polarization effects under the influence of the electric field. In contrast, the ASFA method incorporates the polarization effect by considering orbital distortion. The orbital distortion causes the ionized free-electron wavepacket to exhibit pronounced directionality, thereby enhancing the coherence between the generated ionic states. To illustrate this point, we present the density distributions of the distorted orbitals and the ionized electrons in Fig.~\ref{figure22}. The free-electron wavefunction ionized from the $i$-th MO is calculated by $\psi_{i}^{(\text{free})}(\mathbf{r})\propto\int \mathbf{u}_{i}(\mathbf{k})\!\cdot\! \mathbf{F}e^{i(\mathbf{k}\cdot\mathbf{r})}d\mathbf{k}$. As shown in Figs.~\ref{figure22}(a) and \ref{figure22}(c), the orbitals are stretched along the $-z$ direction under the influence of the electric field. Consequently, these two distorted MOs tend to ionize along the $-z$ direction, see Figs.~\ref{figure22}(b) and \ref{figure22}(d).
This increases the probability that the two ionic states are correlated with the same free electron. As a result, the coherence between the two ionic states is enhanced.
However, the ``Simple" and PWE methods adequately address, and even overestimate, the directionality of strong-field ionization. For instance, the PWE method considers only the ionization from the asymptotic regions of the orbital wavefunctions in the opposite direction of the electric field. Hence, these two methods are better suited for describing the coherence produced by tunnelling ionization, but they may overestimate the coherence generated at low laser intensities.
In contrast, the ASFA method can be applied across a broader range of laser intensities due to its inclusion of field-dependent orbital distortion.

\subsection{Post-ionization dynamics in N$_2^+$} 
Next, we focus on the influence of ionization-produced coherences on the dynamics of N$_2^+$. In the simulations, we use a linearly polarized 30-fs, 800-nm laser pulse with an intensity of $3 \times 10^{14}$ W/cm$^2$. Changing the wavelength will influence the effect of the ionization-produced coherence on the dynamics of N$_2^+$, as reflected in Eq.~(\ref{eq:2state2}), which will be discussed later. Additionally, as discussed in our previous work \cite{xue2021vibronic}, the wavelength determines the pattern of the vibronic-state DOC after the pulse is over, with the maximum DOC occurring at $\omega^{ij}_{vv'}=n\omega$. For ionizing orbitals with the same parities, $n$ is even; for those with opposite parities, $n$ is odd.

\begin{figure} [htb]
\includegraphics[width=8.5cm,clip=true]{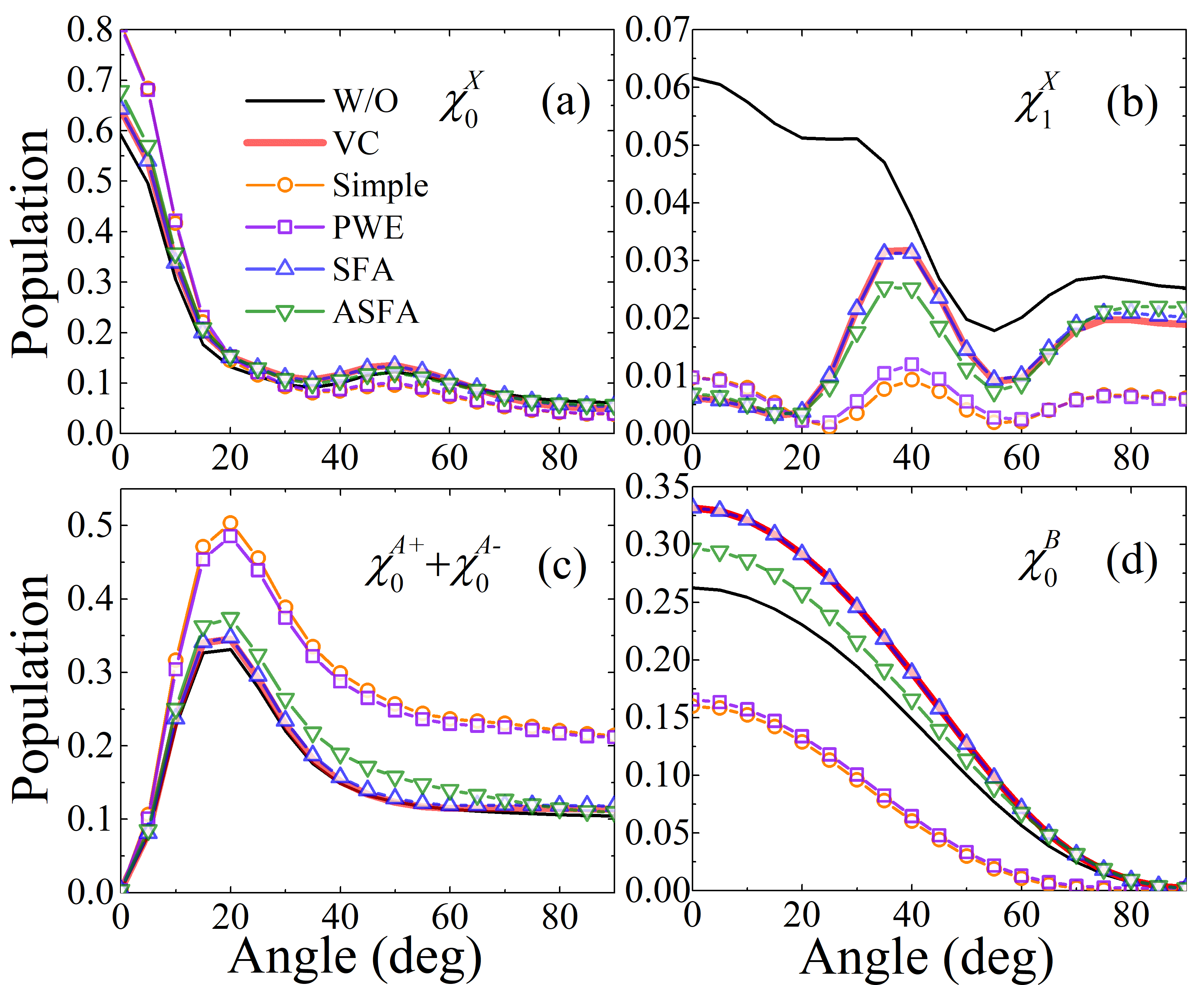}
\caption{(Color online) (a)-(d) Populations of $\chi_0^X$, $\chi_1^X$, $\chi_0^{A_-}+\chi_0^{A_+}$, and $\chi_0^B$, i.e., $\rho_{00}^{XX}$, $\rho_{11}^{XX}$, $\rho_{00}^{A_-A_-}+\rho_{00}^{A_+A_+}$ and $\rho_{00}^{BB}$, as functions of the molecular axis angle. ``W/O" refers to results calculated without considering any ionization-produced coherence. ``VC" denotes results that include only the ionization-produced vibrational coherence.   
``Simple", ``PWE", ``SFA", and ``ASFA" refer to results incorporating both the vibrational and vibronic coherences, with the vibronic coherences calculated using the corresponding coherence methods. Because the ionization rate and vibrational coherence are the same for different coherence methods, ``W/O" and ``VC" are method independent. The results of 90$^\circ-180^\circ$ are mirror-symmetric to those of 0$^\circ-90^\circ$ and are not displayed here.}
\label{figure33}
\end{figure} 

Figure \ref{figure33} shows the populations of $\chi^{X}_{0}$, $\chi^{X}_{1}$, $\chi^{A}_{0}$, and $\chi^{B}_{0}$ as functions of the molecular axis angle calculated under different conditions. These vibronic states are associated with N$_2^+$ lasing at 391 nm and 428 nm \cite{PhysRevA.84.051802,Xu2015,PhysRevLett.115.133203,PhysRevLett.119.203205,PhysRevLett.120.133208,PhysRevLett.129.123002}. 
By comparing the ``VC" results with the ``W/O" results, we find that the ionization-produced VCs cause a significant decrease in the population of $\chi_1^X$, and an increase in the population of $\chi_0^B$ by up to $\sim 25\%$. In contrast, the populations of $\chi_0^X$ and $\chi_0^A$ are almost unaffected by the VCs. 
To access the effect of ionization-produced VECs, we further compare the ``Simple", ``PWE", ``SFA", and ``ASFA" results with the ``VC" results. Several key features can be observed.
(i) The ``SFA" results overlap with the ``VC" results. This is because the SFA predicts the absence of vibronic coherences for $X-A_\pm$ and $X-B$ (see Figs.~\ref{figure11}(a) and \ref{figure11}(b)). 
(ii) Compared with the ``VC" results, the populations calculated by the ``Simple", PWE, and ASFA methods undergo similar changes, although their magnitudes differ.
For $\chi^X_0$ and $\chi^X_1$, the ionization-produced VECs result in an increase in population at small angles. For $\chi^A_0$ and $\chi^B_0$, the VECs induce population increases and decreases at nearly all angles, respectively.
(iii) The ``Simple" results exhibit the largest population changes, followed by the ``PWE" results, while the ``ASFA" results show relatively smaller changes. 
This finding is supported by the DOCs shown in Figs.~\ref{figure11}(a-d), where the ``Simple" method predicts the largest VECs, followed by the ``PWE" results, and the ASFA method predicts smaller VECs.  

To understand the effect of ionization-produced VCs on the transitions between ionic states, we employ a three-state $\Lambda$ (or V)-type model. States 1 and 3 are close in energy, mimicking two adjacent vibrational states of the same electronic state. State 2 mimics a vibronic state of a different electronic state. There are non-zero TDMs from states 1 and 3 to state 2. The quantum Liouville equations for this three-state system are as follows,
\begin{small}
\begin{equation}\label{eq:3state1}
\begin{aligned}
\dot{\rho}_{22}&=-2u_{12}F(t)\text{Im}[\rho_{12}] -2u_{32}F(t)\text{Im}[\rho_{32}],  \\
\dot{\rho}_{12}&=-\text{i}\omega_{12}{\rho}_{12}+ \text{i}u_{12}F(t)(\rho_{22}-\rho_{11}) -\text{i}u_{23}F(t)\rho_{13},
\\
\dot{\rho}_{32}&=-\text{i}\omega_{32}{\rho}_{32}+ \text{i}u_{32}F(t)(\rho_{22}-\rho_{33}) -\text{i}u_{12}F(t)\rho_{13}^*,\\
....
\end{aligned}
\end{equation}
\end{small}%
If the instantaneous ionization-produced coherence $\rho_{13}^{\text{ins}}$ is inserted at $t - \delta t$, the contribution of this term to $\dot{\rho_{22}}(t)$ should be
\begin{equation}\label{eq:3state2}
\dot{\rho}_{22}^{\text{VC}}(t)\approx 4u_{32}u_{12}F^2(t)\delta t \text{Re}[\rho_{13}^{\text{ins}}(t-\delta t)].
\end{equation}%
The sign of $u_{32}u_{12}\times \text{Re}[\rho_{13}^{\text{ins}}]$ determines whether $\rho_{13}^{\text{ins}}$ causes an increase or decrease in $\rho_{22}$. This effect is due to the constructive or destructive interference between the pathways $1\rightarrow 2$ and $3\rightarrow 2$. 
Using Eq.~(\ref{eq:3state2}), we can explain the effect of the ionization-produced VCs. Taking the $X\rightarrow B(v=0)$ transition as an example, we analyze the transition process. For the $X$ electronic state, the most populated vibrational states are $\chi_{0}^X$ and $\chi_{1}^X$. Therefore, the $X\rightarrow B(v=0)$ transition is dominated by the pathways $\chi_{0}^X \rightarrow \chi_{0}^B$ and $\chi_{1}^X \rightarrow \chi_{0}^B$. Specifically, states $\chi_{0}^X$, $\chi_{0}^B$, and $\chi_{1}^X$ correspond to states 1, 2, and 3 in the three-state model, respectively. And $\rho_{01}^{XX,\text{ins}}$ corresponds to $\rho_{13}^{\text{ins}}$.    
Given that $\rho_{01}^{XX,\text{ins}}\propto C_{0}^{X(\text{FC})}C_{1}^{X(\text{FC})}>0$, $u^{XB}_{00}=-0.60~\text{a.u.}<0$ and $u^{XB}_{10}=-0.38~\text{a.u.}<0$, the two pathways interfere constructively. Therefore, when incorporating the ionization-produced VCs, $\rho_{00}^{BB}$ increases appreciably, while $\rho_{11}^{XX}$ decreases significantly. 
The population changes of other vibronic states can also be analyzed in a similar manner.

To understand the effect of the ionization-produced VECs, we adopt a two-state model interacting with an electric field
$F(t)=  F_0  \text{cos}(\omega t+\phi )$.  
Under the rotating wave approximation, the two-state DM equations can be written as
\begin{equation} \label{eq:2state1}
\begin{aligned}
&\dot{\tilde {\rho}}_{12}=\text{i}u_{12}\frac{F_0}{2} e^{-\text{i}\epsilon t+\text{i}\phi }
\Delta+\rho_{12}^{\text{ins}}e^{-\text{i}\omega_{21}t},\\
&\dot{\Delta}=-2u_{12}F_0\text{Im}\left ( \tilde {\rho}_{12}e^{\text{i}\epsilon t-\text{i}\phi } \right ) , 
\end{aligned}
\end{equation}%
where $\Delta=\rho_{22}-\rho_{11}$, $\rho_{12}=\tilde{\rho}_{12}e^{-\text{i}\omega_{12}t}$ and $\epsilon =\omega_{21}-\omega$. 
$\rho_{12}^{\text{ins}}$ represents the instantaneous ionization-produced VEC.
Because $\tilde {\rho}_{12}(t)=\tilde {\rho}_{12}(t-\delta t)+\rho_{12}^{\text{ins}}(t-\delta t) e^{-\text{i}\omega_{21}(t-\delta t)}\delta t+O(\Delta)$, the population transfer caused by $\rho_{12}^{\text{ins}}$ can be assessed by
\begin{equation} \label{eq:2state2} 
\Delta^{\text{VEC}} =-2u_{12}F_0\text{sin}(\epsilon \delta t)\!\!\int\!\!\text{Re}[\rho_{12}^{\text{ins}}(t)]   \text{cos}(\omega t+\phi )  dt.  
\end{equation}%
In the derivation, we consider that the imaginary part of the VEC is negligible (see Appendix).
According to this equation, the population transfer caused by $\rho_{12}^{\text{ins}}$ depends on $u_{12}$, the frequency detuning $\epsilon$ and the integral term.
For the molecular axis angles smaller than $90^\circ$, $\rho_{XB}^{\text{ins}}$ has a sign opposite to the electric field (see Fig.~\ref{figure11}), resulting in a negative integration term. Taking $\chi^X_{0}-\chi^B_0$ and $\chi^X_1-\chi^B_0$ transitions as examples, $\epsilon =\omega_{00}^{BX}(\text{or }\omega_{01}^{BX})-\omega>0$, $u_{00}^{XB}(\text{or }u_{10}^{XB})<0$, so $\Delta^{\text{VEC}}$ is negative. This means that the ionization-produced VEC has a suppressive effect on the transition from $\chi^X_{0}$(or $\chi^X_{1}$) to $\chi^B_0$. Therefore, compared to the ``VC" results, $\rho^{XX}_{00}$ and $\rho^{XX}_{11}$ are increased, while $\rho^{BB}_{00}$ is decreased in the ``Simple", ``PWE", and ``ASFA" results (see Figs.~\ref{figure33}(a), \ref{figure33}(b), and \ref{figure33}(d)). For the $X-A_{\pm}$ transition, since $\omega_{AX}\approx \omega$, the detuning $\omega^{AX}_{v'v}\!-\!\omega$ can be either positive or negative. As a result, the ionization-produced VEC can either promote or suppress the $X\rightarrow A_\pm$ transitions. 
Specifically, for the $\chi^X_{0}-\chi^{A_-}_{0}$ transition, $\Delta^{\text{VEC}}>0$. Namely, the VEC enhances the $\chi^X_{0}\rightarrow\chi^{A_-}_{0}$ transition, thereby resulting in the increase of the $\rho^{A_-A_-}_{00}$ (see Fig.~\ref{figure33}(c)). 
Furthermore, according to Eq.~(\ref{eq:2state2}), we also find that the laser wavelength can be utilized to manipulate the ionization-produced VECs by adjusting the detuning between the vibronic-state energy difference and the laser frequency, thereby controlling the post-ionization molecular dynamics of the residual ions.

\begin{figure} [htb]
\includegraphics[width=9cm,clip=true]{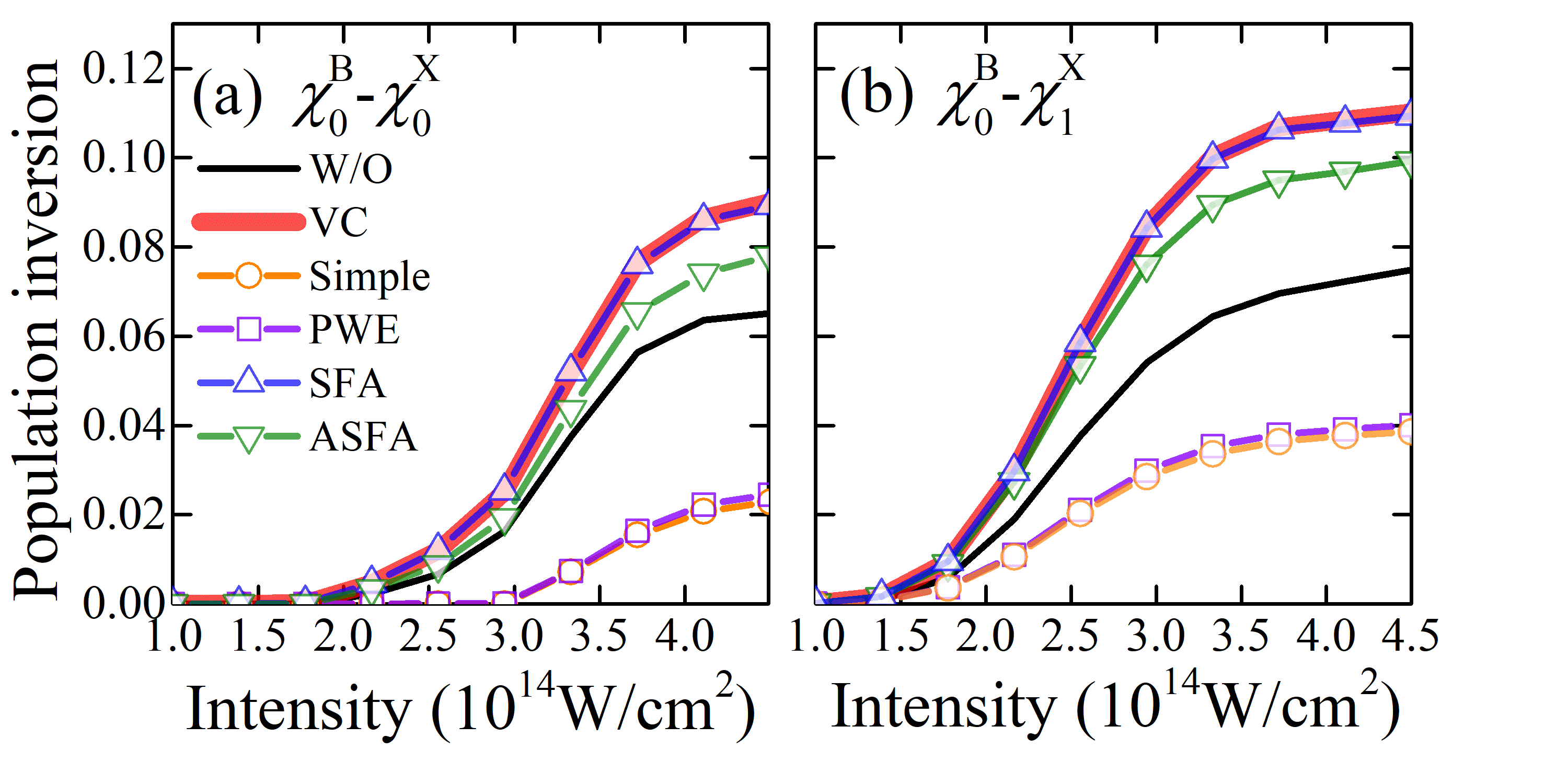}
\caption{(Color online) Angle-integrated population inversions of (a) $\chi_0^X-\chi_0^B$ and (b) $\chi_1^X-\chi_0^B$ as functions of laser intensity. Only angles where the population inversion is greater than zero are included in the integration. The abbreviations used here have the same meanings as those defined in Fig.~\ref{figure33}. 
}
\label{figure44}
\end{figure}

Compared to the populations of individual states, we are more interested in the population inversion between $X$ and $B$, as it directly relates to the N$_2^+$ lasing generation in ambient air \cite{PhysRevA.84.051802,Xu2015,PhysRevLett.115.133203,PhysRevLett.119.203205,PhysRevLett.120.133208,PhysRevLett.129.123002}. Figures~\ref{figure44}(a) and \ref{figure44}(b) respectively show the population inversions between $\chi^X_{0}$ and $\chi^B_0$, and between $\chi^X_{1}$ and $\chi^B_0$ as functions of laser intensity. The results are obtained by integrating the signals over molecular axis angles where the population inversion is positive. Obviously, the population inversions in both cases are strongly enhanced when considering the ionization-produced VCs. In contrast, when the ionization-produced VECs are incorporated, the population inversions are suppressed compared to the VC results, as shown in the ``Simple", ``PWE", and ``ASFA" results. Because the VECs of $X-B$ and $X-A_{\pm}$ predicted by the SFA method are zero, the ``SFA" and ``VC" results nearly overlap.  
Overall, the ``Simple" and ``PWE" results are similar, showing reductions of over 50\% compared to the ``W/O" results. This increases the laser intensity threshold for the $\chi^X_{0}$-$\chi^B_0$ population inversion to approximately $3\times 10^{14}$W/cm$^2$. However, as shown in Fig.~\ref{figure44}(a), the laser intensity threshold predicted by the ASFA method is around $1.5\times 10^{14}$W/cm$^2$, which is consistent with the experimental finding for the 391-nm lasing in Ref.~\cite{liu2015recollision}. 
Moreover, although the results of ASFA are slightly lower than those of the ``VC", ionization-produced coherences still enhance the population inversion of $\chi^X_{0}-\chi^B_0$ by up to $\sim$20\% and that of $\chi^X_{1}-\chi^B_0$ by up to $\sim$30\%.

\subsection{Ionization-produced coherence in O$_2^+$}
We now investigate the post-ionization dynamics of O$_2$, focusing on the influence of the ionization-produced coherence on the dissociation signals of O$_2^+$.
When a strong laser field interacts with O$_2$, electrons can be ionized from the HOMO (1$\pi_{g\pm}$), HOMO-1 (1$\pi_{u\pm}$), and HOMO-2 (3$\sigma_g$) orbitals, resulting in the formation of O$_2^+$ in the X$^2\Pi_{g\mp}$, $a^4\Pi_{u\mp}$ and $b^4\Sigma_g^-$ electronic states, respectively. $\pm$ on the subscript denotes the orbital angular momentum along the molecular axis, with values of $\pm 1$. Subsequently, O$_2^+$ can dissociate along the $f^4\Pi_g$ potential curve by the $a^4\Pi_u \rightarrow f^4\Pi_g$ transition \cite{de2011following,magrakvelidze2012dissociation,corlin2015probing}. Our previous studies verified that the $b^4\Sigma_g^-$ state also plays a crucial role in the dissociation process through a pathway of $b^4\Sigma_g^- \rightarrow a^4\Pi_{u\mp} \rightarrow f^4\Pi_{g\mp}$ when the $b^4\Sigma_g^-$ and $a^4\Pi_{u\mp}$ states are resonantly coupled by the laser field \cite{xue2018following}. However, the $X^2\Pi_{g\mp}$ state is not coupled to the other three states because of its distinct spin multiplicity, and is therefore excluded from the simulations. Consequently, only the states a$^4\Pi_{u\pm}$, $b^4\Sigma_g^-$ and f$^4\Pi_{g\pm}$ are included in the simulations, and denoted as $a_\pm$, $b$, and $f_\pm$ for simplicity. As known, the $a_\pm-f_\pm$ TDMs are parallel to the molecular axis, while the $a_\pm-b$ TDMs are perpendicular to the molecular axis \cite{xue2018following}.

\begin{figure} [htb]
\includegraphics[width=9cm,clip=true]{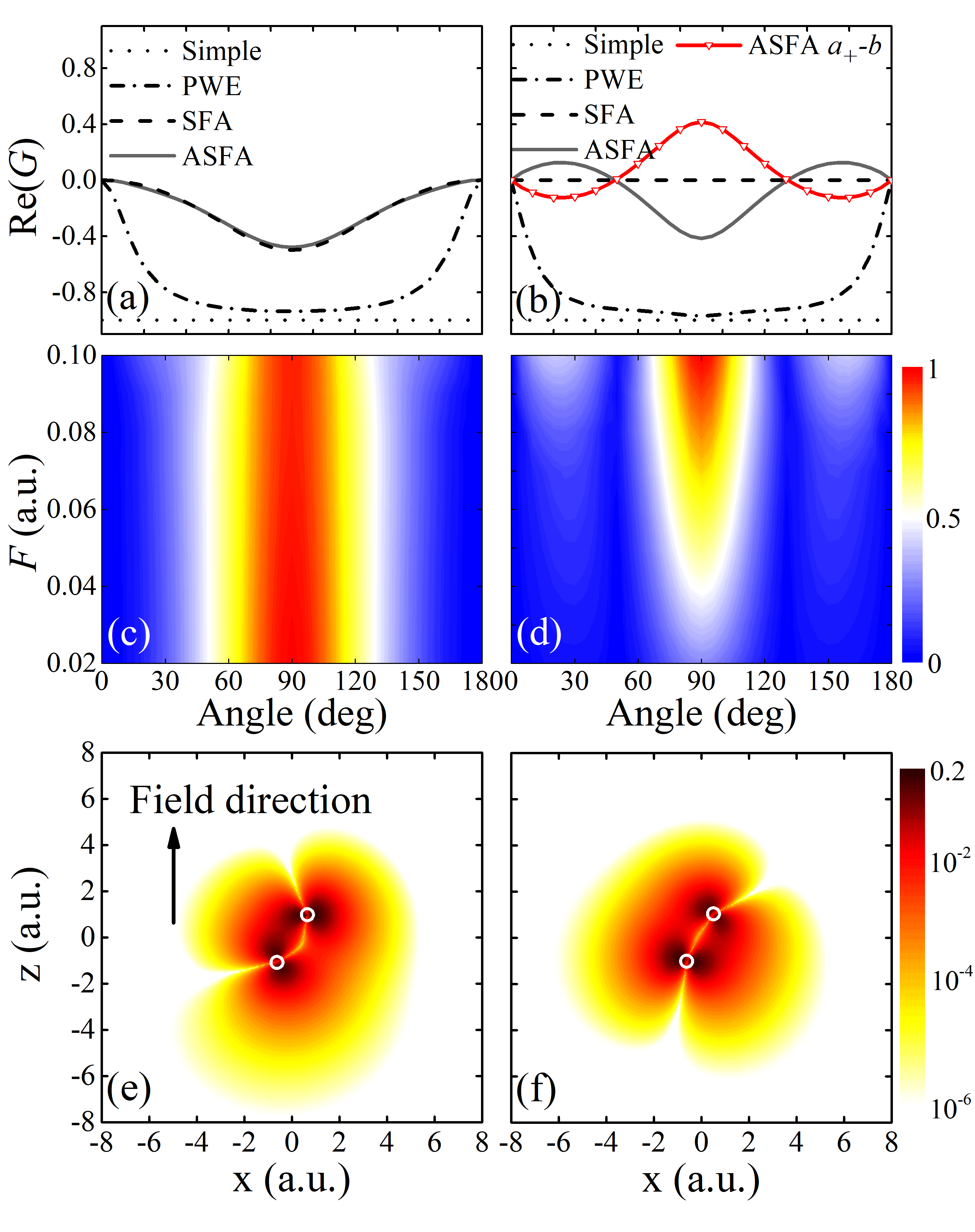}
\caption{(Color online) Real parts of the complex degree of coherences (DOCs) $G_{ij}$ for (a) $a_--a_+$ and (b) $a_--b$ as functions of the molecular axis angle $\beta$, calculated at a field strength of $F=0.09$~a.u. The $a_+-b$ DOC calculated by the ASFA method is also plotted as a red line with triangles. Field-strength-dependent DOC $|G_{ij}|$ for (c) $a_--a_+$ and (d) $a_--b$. Electron density in the $xoz$ plane of the distorted (e) $\uparrow$-spin $1\pi_{u+}$ and (f) $\downarrow$-spin $1\pi_{u+}$ orbitals at $F=0.09$~a.u.~and $\beta=30^\circ$. $\beta$ is the angle between the molecular axis and the field direction.}
\label{figure55}
\end{figure}

First, we discuss the properties of the ionization-produced coherence.   
Figures~\ref{figure55}(a) and \ref{figure55}(b) show the ionization-produced complex DOC $G_{ij}$ as a function of molecular axis angle at a fixed field strength of $F$=0.09 a.u. Similar to Fig.~\ref{figure11}, only the real parts of $G_{ij}$ are displayed.
Three features can be observed.
(i) The coherence between $a_-$ and $b$ calculated by the SFA method is zero, as shown in Fig.~\ref{figure55}(b). This is because the parities of the field-free ionizing orbitals $1\pi_{u\pm}$ and $3\sigma_g$ are opposite, making the corresponding electronic continuum states orthogonal. As a result, there is no coherence between states $a_\pm$ and $b$.  
(ii) Similar to N$_2^+$, the ``Simple" method predicts the highest DOC $|G_{ij}|$, followed by the PWE method, while the SFA and ASFA methods predict lower DOCs $|G_{ij}|$. This feature can be understood as follows. In comparison with the ``Simple" and PWE methods, which primarily consider ionizations occurring in the opposite direction of the field, the SFA (or ASFA) method considers ionizations in all directions. More ionization directions result in a lower probability that the generated ionic states are correlated with the same free electron. Consequently, the DOC between the ionic states reduces.  
(iii) Although the values of Re$(G_{ij})$ predicted by different methods differ, their signs are the same, except for Re$(G_{a_\mp b})$ calculated by the ASFA method around $30^\circ$ and $150^\circ$. This exception can be attributed to the counterintuitive shape of the $1\pi_{u\pm}$ orbital at these angles, from which ionization produces the ionic state $a_\mp$. 
More specifically, in our DFT calculations, the configuration of the ground state X$^3\Sigma_g^-$ in O$_2$ is $KK(2\sigma_g)^2(2\sigma_u)^2(3\sigma_g)^2(1\pi_{u+})^2(1\pi_{u-})^2(1\pi_{g+})^{\uparrow }(1\pi_{g-})^{\uparrow }$, where the two unpaired electrons both have $\uparrow$ spin. Therefore, the $\downarrow$-spin $1\pi_{u\pm}$ orbital should be ionized to produce the spin quartet ionic states $a^4\Pi_{u\mp}$.   
Taking the $1\pi_{u+}$ orbital as an example, we show the electron densities of the distorted orbitals with spin $\uparrow$ and $\downarrow$ at $\beta=30^{\circ}$ in Figs.~\ref{figure55}(e) and \ref{figure55}(f), respectively.    
It can be seen that the $\uparrow$-spin orbital tends to stretch in the opposite direction of the applied field. 
However, the $\downarrow$-spin orbital does not stretch in the opposite direction of the electric field. This is because the $\downarrow$-spin orbital is influenced not only by the field but also by the Coulomb repulsion from the $\uparrow$-spin orbital. This counterintuitive shape of the $\downarrow$-spin $1\pi_{u+}$ orbital ultimately leads to the abnormal sign of the $a_- - b$ coherence around $30^\circ$.

We further discuss the effect of field strength on the DOC. Since the DOCs predicted by the ``Simple", PWE, and SFA methods remain constant or nearly unchanged with varying field strength, we present only the ASFA results in Figs.~\ref{figure55}(c) and \ref{figure55}(d). 
As shown, the $a_+-a_-$ DOC hardly changes with field strength.    
In contrast, the $a_--b$ DOC gradually increases with increasing field strength.   
This indicates that the $a_+ - a_-$ DOC is not sensitive to orbital distortion, whereas the $a_- - b$ DOC is. This difference can be understood as follows.
The ionic states $a_+$ and $a_-$ are generated by the ionization of MOs $1\pi_{u-} $ and $1\pi_{u+}$. The two MOs possess the same parity. 
When not considering orbital distortion, the free-electron wavepackets generated from these two MOs also have the same parity. Consequently, after tracing out the states of the free electron, there is always coherence between states $a_+$ and $a_-$.
In contrast, the ionic state $b$ comes from the ionization of the $3\sigma_g$ orbital, which has the opposite parity to the $1\pi_{u\pm}$ orbitals. 
When not considering orbital distortion, the electrons ionized from the $3\sigma_g$ and $1\pi_{u\pm}$ orbitals can not occupy the same quantum state. Therefore, there is no coherence between states $a_\pm$ and $b$ in the case. Because the ASFA method considers the effect of orbital distortion, the electrons ionized from the  distorted $3\sigma_g$ and $1\pi_{u\pm}$ orbitals can have a nonzero probability of occupying the same quantum state. 
As a result, the $a_\pm - b$ coherence emerges. 
Moreover, the stronger the field strength, the more severe the orbital distortion. Therefore, the coherence between states $a_\pm$ and $b$ increases with increasing field strength.

\subsection{Post-ionization dynamics in O$_2^+$} 
\begin{figure} [htb]
\includegraphics[width=9cm,clip=true]{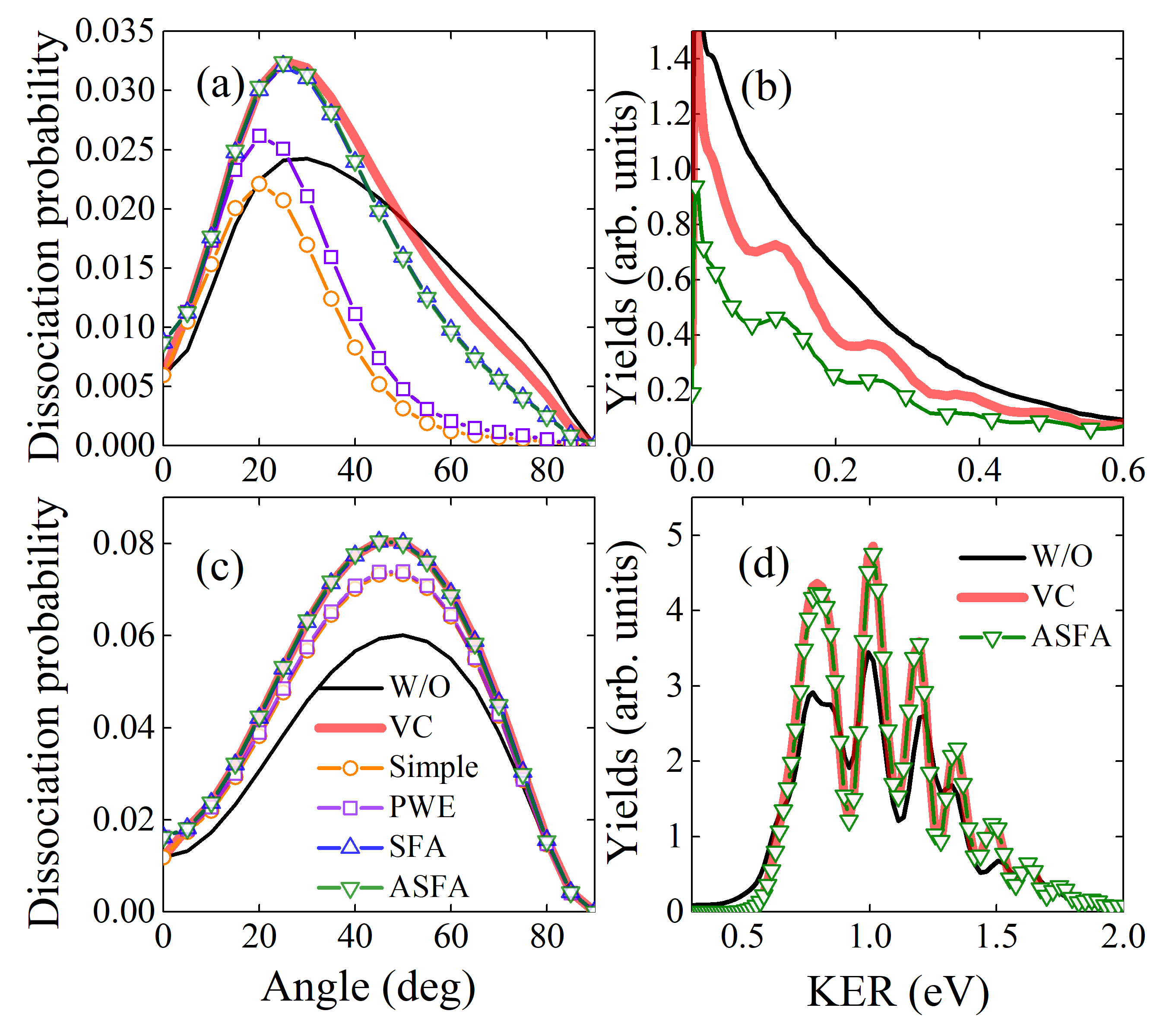}
\caption{(Color online) (a) Dissociation probabilities as functions of the molecular axis angle $\beta$. An 800-nm linearly polarized laser pulse with an intensity of $2.5\times 10^{14}~\text{W/cm}^2$ is used. The abbreviations used here have the same meaning as those defined in Fig.~2. (b) Kinetic-energy-release spectra at $\beta=70^\circ$. (c), (d) Same as (a) and (b), but with a wavelength of 400~nm.}
\label{figure66}
\end{figure}

In the following, we explore the role of ionization-produced coherences in the dissociative ionization dynamics of O$_2$. 
In the simulations, a linearly polarized 800-nm, 30-fs laser pulse with an intensity of $2.5\times 10^{14}\text{W/cm}^2$ is used.
At this wavelength, the electronic states $a_\pm$ and $b$ are resonantly coupled. 
Figure \ref{figure66}(a) shows the dissociation probabilities as a function of the molecular axis angle $\beta$, calculated under different conditions.
By comparing the ``VC" results with the ``W/O" results, it is found that the VC enhances the dissociation probability at small angles while reducing it at large angles. 
This can be explained as follows. At small angles, the $a_{\pm}\rightarrow f_{\pm}$ parallel transition dominates. Based on Eq.~(\ref{eq:3state2}), the ionization-produced VCs in the $a_{\pm}$ state will increase the $a_{\pm}\rightarrow f_{\pm}$ transitions, leading to an enhancement of the dissociation probability. As the angle increases, the $a_\pm \rightarrow b $ perpendicular transitions gradually become strong. The VCs increase the transition of this pathway, thereby reducing the population of the $a_\pm$ state. As a result, the dissociation probability is weakened.
Next, we discuss the effect of the ionization-produced VECs. As shown, compared to the ``VC" results, the ``Simple", PWE, SFA, and ASFA methods predict weaker dissociation signals. 
This is because the ionization-produced $a_+-a_-$ coherence enhances the $a_\pm \rightarrow b$ transition, thereby weakening the dissociative process. The enhancement of the $a_\pm \rightarrow b$ transition can be interpreted within the frame of the $\Lambda $-type three-state model. Specifically, states $a_+$ and $a_-$ correspond to states 1 and 3 in Eq.~(\ref{eq:3state2}), respectively. It can be seen from Fig.~\ref{figure55}(a) that the $a_+-a_-$ coherence is smaller than zero. Moreover, the interacting term satisfies $\mathbf{u}_{a_+b}\cdot \mathbf{F}=-\mathbf{u}_{a_-b}\cdot\mathbf{F}$, which holds for all molecular axis angles and field directions. Consequently, the $a_+\rightarrow b$ and $a_-\rightarrow b$ transitions interfere constructively, leading to an overall enhancement of the $a_\pm \rightarrow b$ transitions.  
In addition, since the SFA and ASFA methods predict comparable $a_+-a_-$ coherence (see Fig.~\ref{figure66}(c)), their predicted dissociation probabilities are nearly identical. However, the ``Simple" and PWE methods predict higher $a_+-a_-$ coherences, so they predict much weaker dissociation signals.

The influence of the ionization-produced coherences is also reflected in the KER spectrum. 
Figure~\ref{figure66}(b) shows the KER spectra calculated at $\beta=70^\circ$. This angle is chosen because ion signals were collected along the direction perpendicular to the laser polarization in previous experiments \cite{zohrabi2011vibrationally,de2011following}. Compared with the ``W/O" result, the intensity of the ``VC" spectrum reduces. After introducing the ionization-produced VEC, the ``ASFA" signals show an additional 40\% decrease.
Moreover, both the ``VC" and ``ASFA" spectra exhibit peak structures, as observed in the single-pulse and the IR-pump-IR-probe experiments \cite{zohrabi2011vibrationally,de2011following}. The peaks are primarily induced by the ionization-produced VC and correspond to the projection of vibrational states $\chi^a_{v>10}$ onto the dissociative continuum states \cite{xue2018following}.
These significant changes suggest the importance of the ionization-produced VCs and VECs in simulating the dissociative ionization dynamics of O$_2$.

To further assess the role of the ionization-produced VECs, we change the laser wavelength to 400 nm. In this case, the resonant transitions between states $a_{\pm}$ and state $b$ are almost closed.
Figure~\ref{figure66}(c) presents the corresponding angle-dependent dissociation probabilities. One can see that only the ionization-produced VCs increase the dissociation signal at nearly all angles by enhancing the $a_{\pm}\rightarrow f_{\pm}$ transitions. However, the influence of the ionization-produced VECs remains notably weak, even for the ``Simple" and PWE methods, where the VECs are expected to be stronger (see Figs.~\ref{figure55}(a) and \ref{figure55}(b)).
Figure~\ref{figure66}(d) shows the corresponding KER spectra at $\beta=70^\circ$. The spectra calculated under the ``VC" condition and by the ASFA method almost overlap and are stronger than the ``W/O" result.
This indicates that the ionization-produced VECs play a negligible role when the electronic states are not strongly coupled.

\begin{figure} [htb]
\includegraphics[width=7cm,clip=true]{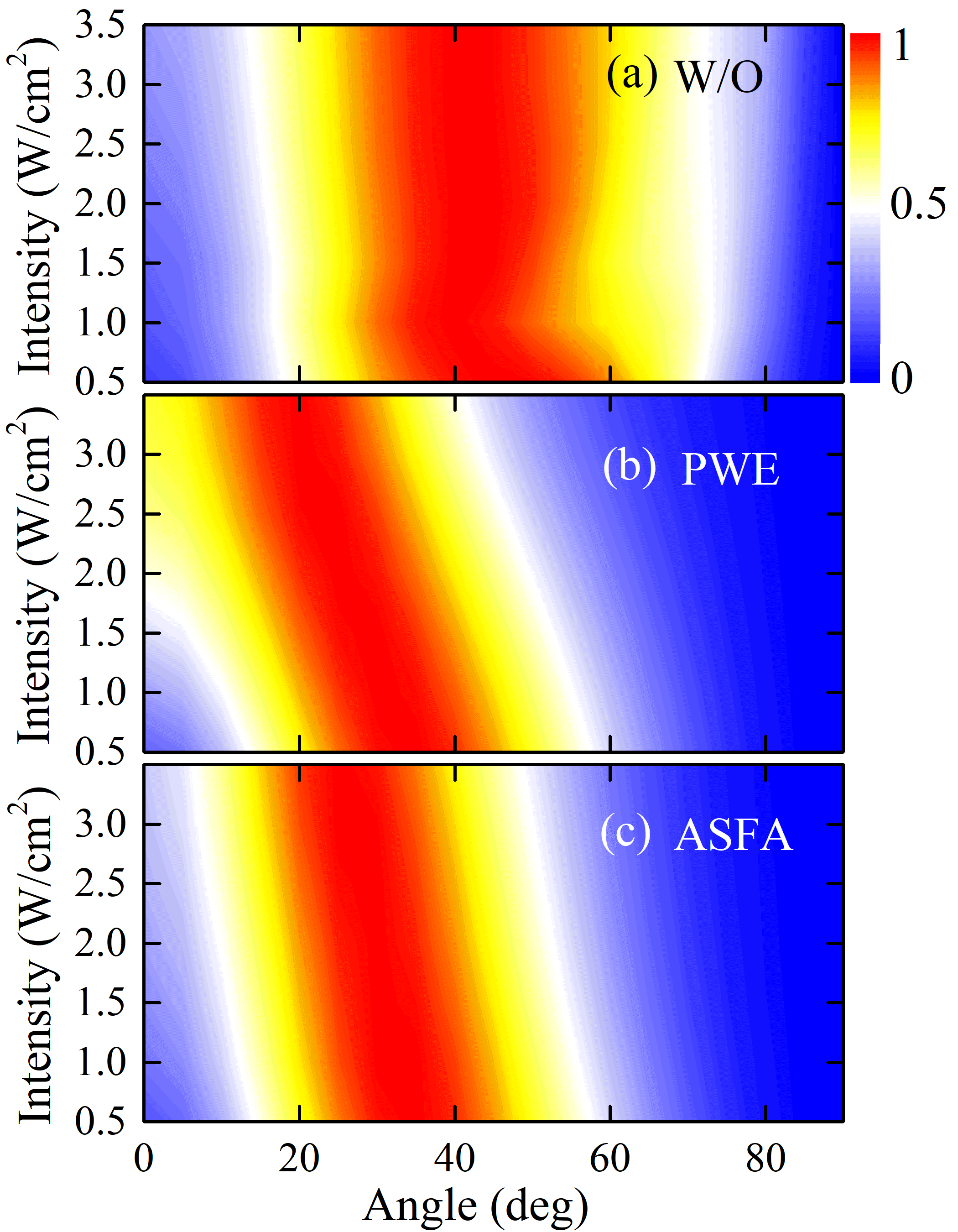}
\caption{(Color online) Normalized angle-dependent dissociation probabilities as functions of laser intensity calculated (a) without considering the ionization-produced coherence, (b) by using the PWE method, and (c) by using the ASFA method, respectively.}
\label{figure77}
\end{figure}%

Finally, in order to detect the signatures of the ionization-produced coherences in O$_2^+$, we suggest performing a laser intensity scan on the angular-dependent dissociation signals. 
Figure \ref{figure77} presents the corresponding results of different methods. An 800-nm 10-fs laser pulse is used here, which can lead to the $a_\pm \rightarrow b$ resonant transition. We focus on the angle with the maximum dissociation signal, which is denoted as the peak angle. As shown, different methods give different slopes for the peak angle as a function of laser intensity.
This phenomenon can be understood by considering the competition between the $a_\pm \rightarrow f_{\pm}$ parallel transition and the $a_\pm \rightarrow b$ perpendicular transition. 
Recalling that the ionization-induced $a_+ - a_-$ coherence enhances the $a_\pm \rightarrow b$ transition, we note two key factors. First, according to Eq.~(\ref{eq:3state2}), the enhancement becomes more pronounced with increasing field strength. Second, the $a_+ - a_-$ DOC increases as the molecular axis angle changes from $0^\circ$ to $90^\circ$ (see Fig.~\ref{figure55}(c)), meaning that the enhancement becomes stronger as the angle approaches $90^\circ$. These two factors lead to a reduced population of the $a_\pm$ state at larger angles with increasing laser intensity, resulting in less dissociation through the $a_\pm \rightarrow f_{\pm}$ pathway. Consequently, the peak angle shifts to a smaller angle with increasing laser intensity.
In the PWE method, the $a_+ - a_-$ coherence is strong (see Fig.~\ref{figure55}(a)), so the signal peak shifts toward smaller angles as the laser intensity increases, as shown in Fig.~\ref{figure77}(b). In the ASFA method, the $a_+ - a_-$ coherence is relatively weak, so the peak angle changes less, as shown in Fig.~\ref{figure77}(c). Under the ``W/O" condition, the ionization-produced coherence is not considered, so the peak angle remains almost unchanged, as shown in Fig.~\ref{figure77}(a). By calibrating the slope of the peak angle, the signatures of the ionization-produced coherence can be detected.

\section{CONCLUSION}
In summary, we present an ASFA method to predict the ionic coherence generated by multi-orbital strong-field ionizations.  
Due to consideration of orbital distortion, the ASFA method overcomes the limitations of the standard SFA, and enables ionic states of different parities to correlate with the same electronic continuum state, resulting in the generation of coherence between these states in the residual ions.
Moreover, we find that both the ``Simple" and PWE methods are more applicable in the tunnelling regime due to the assumption that ionization primarily occurs in the opposite direction of the electric field.
In contrast, the ASFA method can provide reasonable ionization-produced coherence within a broader range of laser intensities.

Besides, we also examine the effect of ionization-produced coherences on the transitions between electronic states in the residual ions.  
For N$_2^+$, the VCs enhance the $X-B$ population inversion, whereas the VECs reduce it. Overall, the ASFA method predicts that ionization-produced coherences increase the population inversion by approximately 25\%$\sim$30\%. As a result, the laser intensity threshold for achieving this inversion is reduced, making it more consistent with the experimental value.
For O$_2^+$, the coherences enhance both the $a_\pm\rightarrow b$ and $a_\pm\rightarrow f_\pm$ transitions. The competition between the two pathways weakens the dissociation via the $a_\pm \rightarrow f_\pm$ transition. By choosing a 400-nm wavelength that nearly closes the $a_\pm \rightarrow b$ transition, we found that the ionization-produced coherences can enhance the dissociation probability by more than 30\%. 
These results highlight the importance of ionization-produced coherences in post-ionization molecular dynamics. 
Accurately manipulating these coherences can offer a promising avenue for controlling post-ionization molecular dynamics and should attract more attention from the strong-field community.

\section*{ACKNOWLEDGEMENT}
This work was supported by the National Natural Science Foundation of China (Grants No.~12274188, No.~12004147, No.~12204209), the Natural Science Foundation of Gansu Province (Grant No.~23JRRA1090), and the Fundamental Research Funds for the Central Universities (Grant No.~lzujbky-2023-ey08).

\section*{APPENDIX: analysis of the coherence phase upon strong-field ionization}
\begin{figure*}[htbp]
\centering
\includegraphics[width=14cm]{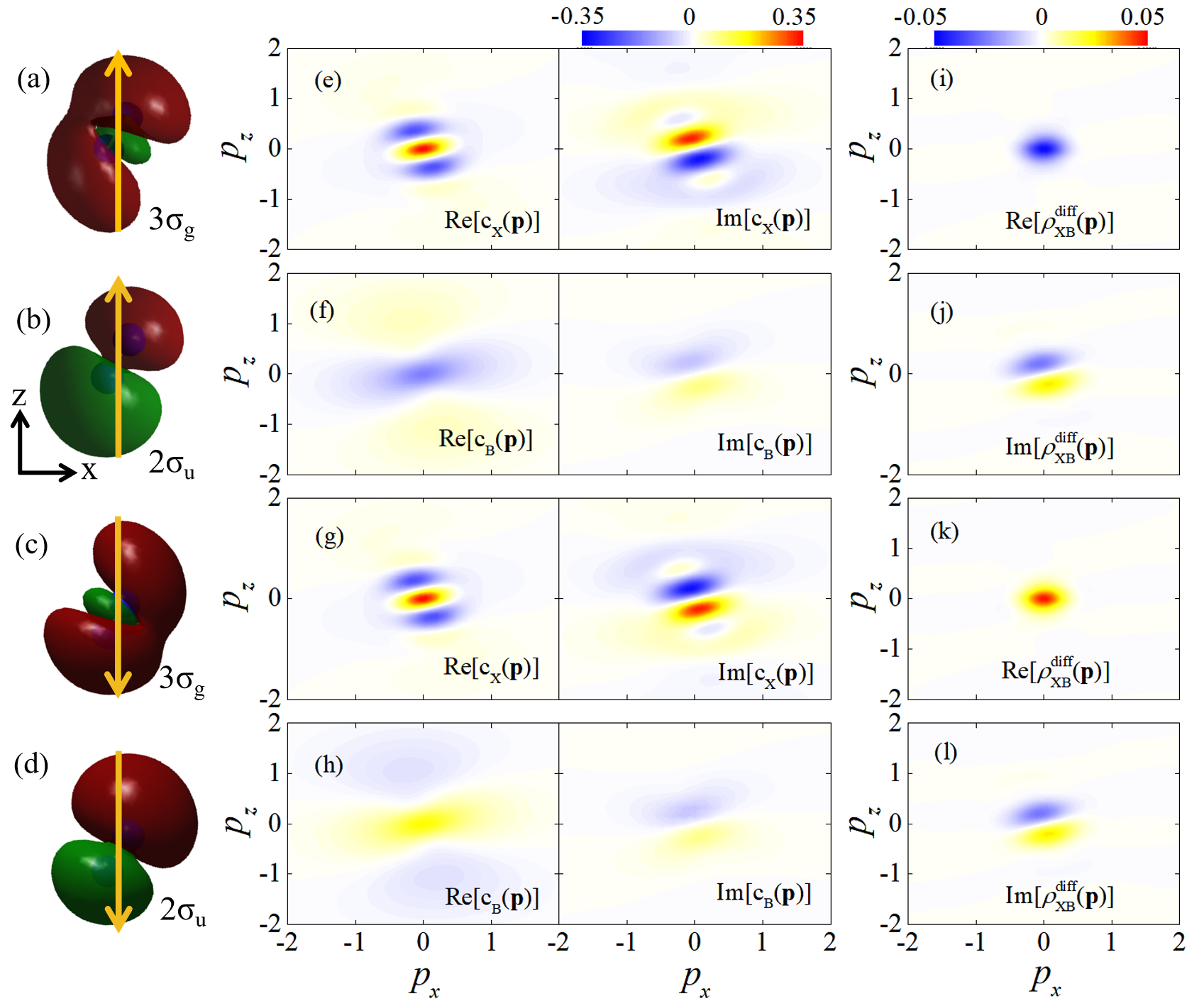}
\caption{(Color online) Distorted (a) HOMO ($3\sigma_g$) and (b) HOMO-2 ($2\sigma_u$) molecular orbitals of N$_2$ at $F=0.08$~a.u.~and $\beta=30^{\circ}$. $\beta$ is the angle between the molecular axis and the space-fixed $z$ direction. (c), (d) Same as (a) and (b), but for $F=-0.08$~a.u. (e)-(h) Real and imaginary parts of the ionization amplitude $c_i(\mathbf{k})$ for the orbitals shown in (a)-(d). (i), (j) Real and imaginary parts of the differential coherence $\rho_{XB}^{\text{diff}}(\mathbf{k})$ at $F=0.08$~a.u. (k), (l) Same as (i) and (j) but for $F=-0.08$~a.u.}   
\label{figure88}
\end{figure*}

Here, we analyze the phase of the ionization-produced coherence calculated by the ASFA coherence method.
In this method, according to Eq.~(\ref{eq:ASFA4}), the instantaneously produced coherence $\rho^{\text{ins}}_{ij}$ depends on $\mathbf{u}_{i}(\mathbf{k})$, where $\mathbf{u}_{i}(\mathbf{k})\!\sim\!-\int  \varphi_i \mathbf{r} e^{-\text{i}\mathbf{k}\cdot\mathbf{r}}d\mathbf{r}$, with $\varphi_i$ representing the wavefunction of the ionizing MO. Due to field-induced distortion, the parity of $\varphi_i$ is destroyed, so that $\mathbf{u}_{i}(\mathbf{k})$ is neither purely real nor purely imaginary. As a result, $\rho^{\text{ins}}_{ij}$ is theoretically expected to be a complex number. However, numerical calculations indicate that $\rho^{\text{ins}}_{ij}$ is still real, which is not easy to understand intuitively. Therefore, a detailed analysis is provided below.

To illustrate this issue, we take the ionization-produced $X-B$ coherence in N$_2^+$ as an example. Figures \ref{figure88}(a)-\ref{figure88}(d) display the distorted HOMO ($3\sigma_g$) and HOMO-2 ($2\sigma_u$) orbitals at external field strengths of $F=\pm 0.08$~a.u. The angle between the molecular axis and the laser polarization ($z$-axis) is set to $\beta=30^\circ$. 
Figures \ref{figure88}(e)-\ref{figure88}(h) show the corresponding ionization amplitude $c_i(\mathbf{k})=\mathbf{u}_i(\mathbf{k})\cdot \mathbf{F}$. Here, $\mathbf{F}=F\mathbf{e}_z$ is the electric field along the space $z$ axis. For simplicity, $\mathbf{k}$ is restricted to the $xoz$ plane. 
It can be seen that Re$(c_i(\mathbf{k}))$ is symmetric with respect to the inversion of $\mathbf{k}$, while Im$(c_i(\mathbf{k}))$ is antisymmetric. We define $\rho_{ij}^{\text{diff}}(\mathbf{k})=c_X(\mathbf{k}) c^*_B(\mathbf{k})$ as the ionization-produced differential coherence. Real and imaginary parts of $\rho_{ij}^{\text{diff}}(\mathbf{k})$ are presented in Figs.~\ref{figure88}(i) and \ref{figure88}(j) for $F=0.08$~a.u., and in Figs.~\ref{figure88}(k) and \ref{figure88}(l) for $F=-0.08$~a.u. As shown, the real part of $\rho_{ij}^{\text{diff}}(\mathbf{k})$ is also symmetric with respect to the inversion of $\mathbf{k}$, while the imaginary part is antisymmetric. 
Recalling that $\rho_{XB}^{\text{ins}}=\int \rho_{ij}^{\text{diff}}(\mathbf{k}) d\mathbf{k}$, the imaginary part cancels out upon integration over $\mathbf{k}$ due to the symmetry properties of $\rho_{ij}^{\text{diff}}(\mathbf{k})$. 
Therefore, $\rho_{XB}^{\text{ins}}$ is real.

Additionally, we also find that the real parts of $\rho_{ij}^{\text{diff}}(\mathbf{k})$ are negative for $F\!>\!0$ and positive for $F\!<\!0$. This behavior can be attributed to the symmetric relation between the MOs produced by the positive and negative fields, i.e.,
\begin{equation}
\begin{aligned}
3\sigma_g(\mathbf{r};F\!>\!0)&=3\sigma_g(-\mathbf{r};F\!<\!0),\\
2\sigma_u(\mathbf{r};F\!>\!0)&=-2\sigma_u(-\mathbf{r};F\!<\!0).
\end{aligned}
\nonumber
\end{equation}%
Therefore, 
\begin{equation}
\begin{aligned}
c_{X}(\mathbf{k};F\!>\!0)&=c_{X}(-\mathbf{k};F\!<\!0),\\
c_{B}(\mathbf{k};F\!>\!0)&=-c_{B}(-\mathbf{k};F\!\!<\!\!0).
\end{aligned}
\nonumber
\end{equation}%
These relations are clearly illustrated in Figs.~\ref{figure88}(a-h).
Finally, one can obtain $\rho^{\text{diff}}_{XB}(\mathbf{k};F>0)=-\rho^{\text{diff}*}_{XB}(\mathbf{k};F<0)$.
As a result, the real part of the coherence changes sign when the field reverses. 

Besides, it is noteworthy that the $3\sigma_g$ orbital exhibits a counterintuitive distortion. Namely, a portion of the red part of the orbital twists along the field direction. This effect becomes more pronounced as $\beta$ approaches 90$^\circ$. This anomalous distortion, likely arising from multi-orbital interactions in the self-consistent field calculation, is closely related to the structure between 60$^\circ$ and 120$^\circ$ in the ASFA's results, as shown in Fig.~\ref{figure11}(a).

%

\bibliography{reference}

\end{document}